\documentclass[a4paper,10pt]{article}
\usepackage{amsfonts}
\usepackage{graphicx}
\usepackage{amsmath}
\usepackage{amssymb}
\usepackage[breaklinks=true,dvips]{hyperref}
\usepackage{breakurl}
\usepackage[utf8x]{inputenc}
\usepackage{mathrsfs}
\usepackage{color}

\title{\bf Optical and spin-optical superpositions modulated by Aharonov-Bohm effect}
\author{Thiago Prud\^encio\footnote{emails: thprudencio@gmail.com; prudencio.thiago@ufma.br} \\ 
\small
\it Coordination of Science \& Technology (CCCT/BICT)
\\ 
\it \small  Federal University of Maranh\~ao (UFMA), \\ 
\small \it 
65080-805, S\~ao Lu\'is-MA, Brazil.}

\date{}

\begin{document}

\maketitle

\begin{abstract}
Generation of Aharonov-Bohm (AB) phase has achieved a state-of-the-art in mesoscopic systems with manipulation and control of the AB effect. The 
possibility of transfer information enconded in such systems to non-classical states of light increases the possible scenarios where 
the information can be manipulated and transfered. In this paper we propose a quantum transfer of the 
AB phase generated in a spintronic device, a topological spin transistor (TST), to an quantum optical device, a coherent state superposition 
in high-Q cavity and discuss optical and spin-optical superpositions in the presence of an AB phase. We demonstrate that the AB phase generated in the TST can be transfered to the coherent state superposition, considering the interaction with the spin state and the quantum optical manipulation of the coherent state superposition. We show that these cases provide examples of two-qubit states modulated by AB effect and that the phase parameter can be used to control the degree of rotation of the qubit state. We also show under a measurement on the spin basis, an optical one-qubit state that can be modulated by the AB effect. In these cases, we consider a dispesive interaction between a coherent state and a spin state with an acquired AB phase and also discuss a dissipative case where a given Lindblad equation is achieved and solved. 
\end{abstract}
{\small
\tableofcontents
}

\section{Introduction}

A typical ilustration of the Aharonov-Bohm (AB) effect \cite{aharo1} is the imprinting of a phase factor 
on the electron's wave function in the region outside a solenoid under a magnetic field applied only in its interior. Another important feature of this effect is that this phase factor, 
the so-called AB phase, generated in the presence of a non-zero gauge potential 
in a region with a null electromagnectic 
field modifies the corresponding energy spectrum of the charged particle \cite{aharo2}. The presence of the AB phase represents a direct influence of a vector potential on the particle dynamics and one of 
remarkable possibilities of quantum interference \cite{olariu3}. 

The emergence of the AB effect in the problem of a charged particle minimally coupled to a gauge field $A_{\mu}$, ($\hbar=1$)
$\hat{H}=\frac{\left({\bf p}-e{\bf A}\right)^{2}}{2m}$,
where $A_{\mu}=(0,{\bf A})$, for a singly connected region, with ${\bf B}=\nabla\times {\bf A}=0$, 
has as a consequence 
the splitting of the wave function in two parties with the acquired AB phase term \cite{caprez4,hernandez5}. Considering a magnetostatic field,  
the whole path closed and the gauge fixing contribution of the AB phase shift term is given by $-i\phi_{AB}= -ie\oint {\bf A}\cdot d{\bf l}$. The interference in the electron states coming 
from the right and left parts around the solenoid give rise to a superposition state with the acquired AB phase given by
$|\psi\rangle =|\downarrow\rangle + e^{i\phi_{AB}}|\uparrow\rangle.$
This  can be undertaken from an interaction between a source-field and the electron, as discussed in \cite{vaidman6}, with the whole 
state given by 
$|\chi\rangle_{I}=\frac{1}{\sqrt{2}}\left(|\uparrow\rangle_{e}|\psi_{L}\rangle_{S} + |\downarrow\rangle_{e}|\psi_{R}\rangle_{S} \right)$, 
colapsing to a final configuration where the left and right states are identical except for the presence of the AB phase,
$|\psi_{L}\rangle_{S}=|\psi\rangle_{S}$ and $|\psi_{R}\rangle_{S}=e^{i\phi_{AB}}|\psi\rangle_{S}$, leaving the total 
state in the following final form  
$|\chi\rangle_{F}=\frac{1}{\sqrt{2}}|\psi\rangle_{S}\left(|\uparrow\rangle + e^{i\phi_{AB}}|\downarrow\rangle\right)$. 

Methods of generating AB phases have growth in different fields, representing 
an important role in solid-state interferometers \cite{duca,niu}, nanotubes \cite{charlier}, transmission microscopy \cite{edgcombe}, 
AlSb/InAs heterostructures \cite{nitta}, quantum Hall effect regime \cite{splet}, Kondo resonance \cite{eckle} and spin transport \cite{matityahu}. Detection of this 
effect in photonic states \cite{li}, optical induction in mesoscopic systems \cite{sigurdsson} 
and interaction of AB ring in a high-Q cavity \cite{alexeev,alexeev2,ramaglia} are examples of the increasing importance of 
such effect in interacting light-and-matter systems and optical analogues of the AB effect \cite{vieira}. 

Advances in the domain of artificial gauge fields have allowed new possibilities in the application of AB effect and geometrical phases \cite{dalibard}.
Optical Berry phase \cite{mironova}, Aharonov-Casher effect \cite{casher,andrade2} as well as the AB effect for neutral particles 
have also been explored in several contexts \cite{kovalev,andrade3,bakke,hartnoll}. The possibility of controlling the AB effect in different scenarios and 
retransmit the phase from the domain where the effect was originally created is an important step in the phenomenological control and 
the manipulation of the associated information. 
Such a realization is particularly important for quantum information and computation tasks. Gate controlled AB oscillations \cite{ding} and 
topological spin transistor (TST) \cite{maciejko} are advances in such direction. The last one makes use of quantum spin Hall insulators, 
experimentally realized in HgTe quantum wells \cite{brune}, with particular importance in spintronics applications \cite{zutic}. Analog of a Datta-Das spin transistor \cite{das}, in the TST the properties of the quantum spin Hall insulator (QSHI) are used to generate an AB phase, that makes and an 
effective rotation of the outgoing spin state with respect 
to the ingoing spin \cite{maciejko}. The control of AB phases \cite{liberto} can also be used for bit codification of the information associated to the AB effect. In the case of a TST, the presence or absence of flux is associated to the presence or not of a phase in the outgoing electron state. 

Here, we discuss optical and spin-optical superpositions in carrying an AB phase. We start considering the instance of a TST as spintronic device to implement a quantum transfer to a coherent state in a high-Q cavity as quantum optical device. Furthermore, we show that manipulation and control of the AB effect in a TST allows the generation of spin superpositions with an AB phase while the interaction with a coherent state in a high-Q cavity leads to the possibility of 
manipulating this phase in the quantum optical scenario. We consider a setup consisting of a TST coupled to a high-Q cavity carrying initially a non-classical state of light, a coherent state, and a detector that project the 
final spin state in one specific state (see figure \ref{cavity}). Coupling a quantum ring to high-Q cavities have been considered previously in other contexts \cite{alexeev}. The high-Q cavity is built 
experimentally using supermirrors with extremelly low losses, achiving Q factor of the order $10^{11}$. The integration with 
a TST can be realized in the implementation with for instance a photonic cristal cavity \cite{muller}.The AB phase generated in the TST can be quantum transfered to the coherent state superposition, considering the interaction with the spin state 
and the quantum optical manipulation of the coherent state superposition. We demonstrate that the AB phase generated in the TST can be transfered to the coherent state superposition, considering the interaction with the spin state and the quantum optical manipulation of the coherent state superposition. We will show that these cases provide examples of two-qubit states modulated by AB effect and the phase parameter can be used to control the degree of rotation of the qubit state. 
We also discuss the fact that a flux control can then be associated to a binary codification, allowing a sequence (string) of bits of a given length be generated in such a way. In particular, the AB phase can be quantum transfered from the spintronic device to a quantum optical one, increasing the possible scenarios where the information can be manipulated and stored. Finally, We will discuss a particular dissipative scenario, by means which we provide and solve a given Lindblad equation for coupling the coherent state and a spin state with an acquired AB phase to a thermal revervoir. 

This work is organized as follows: In Sec. II, we discuss the generation and bit-enconding of AB phases by exploring the AB effect. In 
Sec III, we propose the quantum transfer of AB phase to a coherent state superposition. In Sec. IV, we discuss the use of quantum transfer to 
store a string of bits enconded by AB phases. Sec. V is reserved to our concluding remarks.

\section{TST and coupling to high-Q cavity device}

\begin{figure}
\centering
\includegraphics[scale=0.4]{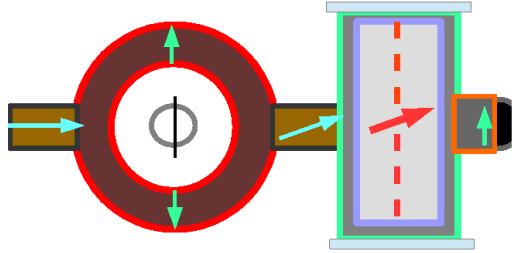}
\caption{(Color online) Theoretical scheme of a TST coupled to a high-Q cavity.  The electron is injected in the $x$-direction and splits in the 
QSH ring to the edge states in up and down direction. Due to a quantum flux, an Aharonov-Bohm phase is acquired by the outgoing state that interacts with a high-Q cavity state (dashed 
red lines) and finally an electron detection is realized in a specific direction.}
\label{cavity1}
\end{figure}

When the outgoing spin state is injected dispersivelly in the high-Q cavity, the interaction between the coherent state and the spin state, in a dispersive regime, 
can generate an entanglement. The setup starts with the electron in the ferromagnet lead being injected into QSHI ring in a $x$-spin polarized 
direction, $|\rightarrow\rangle=|\downarrow\rangle + |\uparrow \rangle$.
Due the properties of the QSHI, the state splits in the ring setup according to the helical properties of the edge states, the edge states of 
opposite spin polarization counterpropagate in the of spin up
($z$ clockwise direction) $|\uparrow\rangle $ and spin down ($-z$ counterclockwise direction) $|\downarrow\rangle$, in the top and bottom edges of the 
ring, respectivelly \cite{maciejko}. The presence of a magnetic flux into the middle of 
the ring inserts a AB phase $\phi_{AB}=2\pi \phi/\phi_{0}$, with $\phi_{0}=hc/e$, in the outgoing electron. This electron state is rotated 
in the $xy$ plane leading to a state of the form
\begin{eqnarray}
|\psi\rangle =|\downarrow\rangle + e^{i\phi_{AB}}|\uparrow\rangle. \label{e}
\end{eqnarray} 
When the outgoing state is injected in a high-Q cavity where there is a defined coherent state $|\alpha\rangle$, with a dispersive interaction of the outgoing state with the coherent state in an 
off-resonant regime, the coupling interaction in the high-Q cavity can be written by
\begin{eqnarray}
\hat{H}_{int}= \beta\hat{n}\otimes\hat{\xi},
\label{ef1}
\end{eqnarray}
where $\beta$ is an effective coupling and the number operator $\hat{n}=\hat{a}^{\dagger}\hat{a}$ is associated to the 
quantized electromagnectic field, with $\hat{a}$ and $\hat{a}^{\dagger}$ the creation 
and annihilation operators acting on the coherent state and the single-electron operator given by
\begin{eqnarray}
\hat{\xi}=\sum_{i,j=\uparrow,\downarrow}|i\rangle\langle j|.
\end{eqnarray}
The quantized electromagnetic 
field in a high-Q cavity initially in a coherent state $|\alpha\rangle$ interacts effectivelly with the 
electron states $|\uparrow\rangle$ and $|\downarrow\rangle$, where a degeneracy $\omega_{e}$ is assumed, corresponding to electron states of same 
kinectic energy. The single-mode field initially in the coherent state $|\alpha\rangle$ has an associated frequency 
$\omega_{\alpha}$, such that a energy difference between the electron kinectic energy and the single-mode frequency is given  
$\delta = |\omega_{\alpha}-\omega_{e}|$. A prototipical ilustration of such scheme is given in the figure \ref{cavity1}, where a spin state is injected in the $x$-direction and a superpostion is generated along ring in a superposition of edge states in up and down direction. The AB phase is acquired by the outgoing state that interacts with a high-Q cavity state. 

For large detuning and short evolving times, where the magnetostatic flux is much slower than 
the frequency of the quantized electromagnetic field, the dispersive interaction between the AB-shifted electrons and the single-mode field 
can be considered. In order to see that the interaction (\ref{ef1}) in fact leads to a phase entanglement 
with the coherent state for an appropriate time of interaction, we can act with the Hamiltonian
\begin{eqnarray}
\hat{H}_{int}|\uparrow\alpha\rangle &=& \hat{n}\beta (|\uparrow \alpha\rangle + |\downarrow\alpha\rangle),\\
\hat{H}_{int}|\downarrow \alpha\rangle &=& \hat{n}\beta (|\uparrow \alpha\rangle + |\downarrow\alpha\rangle),\\
\hat{H}_{int}^{2}|\uparrow\alpha\rangle &=& 2(\hat{n}\beta)^{2} (|\uparrow\alpha\rangle + |\downarrow\alpha\rangle),\\
\hat{H}_{int}^{2}|\downarrow\alpha\rangle &=& 2(\hat{n}\beta)^{2} (|\uparrow\alpha\rangle + |\downarrow\alpha\rangle),\\
\hat{H}_{int}^{3}|\uparrow\alpha\rangle &=& 4(\hat{n}\beta)^{3} (|\uparrow\alpha\rangle + |\downarrow\alpha\rangle),\\
\hat{H}_{int}^{3}|\downarrow\alpha\rangle &=& 4(\hat{n}\beta)^{3} (|\uparrow\alpha\rangle + |\downarrow\alpha\rangle).
\end{eqnarray}
In the general case, it follows 
\begin{eqnarray}
\hat{H}_{int}^{k}|\sigma\alpha\rangle = 2^{k-1}(\hat{n}\beta)^{k} (|\uparrow\alpha\rangle + |\downarrow\alpha\rangle), k\geq 1,\sigma=\uparrow,\downarrow 
\end{eqnarray}
The unitary evolutions are then given explicitly by
\begin{eqnarray}
\sum_{k=0}^{\infty}\frac{(-it)^{k}}{k!} \hat{H}_{int}^{k}|\sigma\alpha\rangle &=& |\sigma\alpha\rangle + \sum_{k=1}^{\infty}\frac{(-it)^{k}}{k!}
2^{k-1}(\hat{n}\beta)^{k} (|\uparrow\alpha\rangle + |\downarrow\alpha\rangle), \sigma=\uparrow,\downarrow \nonumber
\end{eqnarray}
the terms $k=0$ implies the presence of a sign contribution.
\begin{eqnarray}
\sum_{k=0}^{\infty}\frac{(-it)^{k}}{k!} \hat{H}_{int}^{k}|\uparrow\alpha\rangle 
&=& \frac{1}{2}e^{-2i\hat{n}\beta t}(|\uparrow\alpha\rangle + |\downarrow\alpha\rangle) + \frac{1}{2}(|\uparrow\alpha\rangle - |\downarrow\alpha\rangle), \\
\sum_{k=0}^{\infty}\frac{(-it)^{k}}{k!} \hat{H}_{int}^{k}|\downarrow\alpha\rangle 
&=& \frac{1}{2}e^{-2i\hat{n}\beta t} (|\uparrow\alpha\rangle + |\downarrow\alpha\rangle) + \frac{1}{2}(-|\uparrow\alpha\rangle + |\downarrow\alpha\rangle).
\end{eqnarray}
As a consequence, we can write the unitary evolutions as
\begin{eqnarray}
e^{-i\hat{H}_{int}t}|\uparrow\alpha\rangle&=&\frac{1}{2}(|e^{-2i\beta t}\alpha\rangle + |\alpha\rangle)|\uparrow\rangle +\frac{1}{2}(|e^{-2i\beta t}\alpha\rangle - |\alpha\rangle)|\downarrow\rangle,
\end{eqnarray}
\begin{eqnarray}
e^{-i\hat{H}_{int}t}|\downarrow\alpha\rangle&=&\frac{1}{2}(|e^{-2i\beta t}\alpha\rangle - |\alpha\rangle)|\uparrow\rangle + \frac{1}{2}(|e^{-2i\beta t}\alpha\rangle + |\alpha\rangle)|\downarrow\rangle.
\end{eqnarray}

\section{Quantum Transfer of the Aharonov-Bohm phase to the coherent state superposition}

In the case of the electron superposition carrying the AB phase $|\psi\rangle = |\downarrow\rangle + e^{i\phi_{AB}}|\uparrow\rangle$
the state after interaction is the following
\begin{eqnarray}
|\psi(t)\rangle &=& \left[\frac{1}{4}\left(1 + e^{i\phi_{AB}}\right)e^{-2i\hat{n}\beta t} -
\frac{1}{4}\left(1 - e^{i\phi_{AB}}\right)\right]|\downarrow\alpha\rangle \nonumber \\
&+& \left[\frac{1}{4}\left(1 + e^{i\phi_{AB}}\right)e^{-2i\hat{n}\beta t} + \frac{1}{4}\left(1 - e^{i\phi_{AB}}\right)\right]|\uparrow\alpha\rangle. 
\label{c123}
\end{eqnarray} 
In order to project the whole state in a non-classical state of light carrying the AB phase, a detector measures the electron state 
in one of the spin polarizations $|\uparrow\rangle$ or $|\downarrow\rangle$.

%
We can rewrite eq. (\ref{c123}) with the terms of AB phase directly in the coherent state superposition, 
\begin{eqnarray}
|\psi(t)\rangle &=& \frac{1}{4}\left[\left(1 + e^{i\phi_{AB}}\right)|e^{-2i\beta t}\alpha\rangle 
-\left(1 - e^{i\phi_{AB}}\right)|\alpha\rangle \right]|\downarrow\rangle \nonumber \\
&+& \frac{1}{4}\left[\left(1 + e^{i\phi_{AB}}\right)|e^{-2i\beta t}\alpha\rangle 
+\left(1 - e^{i\phi_{AB}}\right)|\alpha\rangle \right]|\uparrow\rangle.
\label{23ec}
\end{eqnarray}
This state can also be simplified in the form 
\begin{eqnarray}
|\psi_{\alpha\alpha'}\rangle &=& \left(\frac{1}{4}\left(1 + e^{i\phi_{AB}}\right)|\alpha'\rangle 
-\frac{1}{4}\left(1 - e^{i\phi_{AB}}\right)|\alpha\rangle \right)|\downarrow\rangle \nonumber \\
&+& \left(\frac{1}{4}\left(1 + e^{i\phi_{AB}}\right)|\alpha'\rangle +\frac{1}{4}\left(1 - e^{i\phi_{AB}}\right)|\alpha\rangle \right)|\uparrow\rangle,\label{ifno}
\end{eqnarray} 
where $\alpha'=\alpha'(t)$ is related to $\alpha$ by means of $\alpha'= e^{-2i\beta t}\alpha$.
If after interaction the spin polarization is detected in the state $|\uparrow\rangle$, the non-classical state is projected in the following 
superposition of coherent states carrying the AB phase
\begin{eqnarray}
|\phi_{+}\rangle = \frac{1}{4}\left(1 + e^{i\phi_{AB}}\right)|\alpha'\rangle +\frac{1}{4}\left(1 - e^{i\phi_{AB}}\right)|\alpha\rangle.\label{kl123}
\end{eqnarray}
On the other hand, if the measurement is realized in the spin polarization $|\downarrow\rangle$, we will have 
\begin{eqnarray}
|\phi_{-}\rangle = \frac{1}{4}\left(1 + e^{i\phi_{AB}}\right)|\alpha'\rangle - \frac{1}{4}\left(1 - e^{i\phi_{AB}}\right)|\alpha\rangle.\label{kl124}
\end{eqnarray}
We can specify the time of interaction in $t=\pi/2\beta$, adjusting the high-Q cavity parameters, such that we now have for each case
\begin{eqnarray}
|\phi_{\pm}\rangle = \frac{1}{4}\left(1 + e^{i\phi_{AB}}\right)|-\alpha\rangle \pm \frac{1}{4}\left(1 - e^{i\phi_{AB}}\right)|\alpha\rangle.
\label{fincodm}
\end{eqnarray}
The coefficients in (\ref{fincodm}) can be rearranged by the action of 
a Hadamard-type gate operation in the basis of coherent states \cite{prudencio}, 
\begin{eqnarray}
\hat{U}_{H}&=&\frac{1}{\sqrt{2}}(|-\alpha \rangle\langle -\alpha|-|\alpha \rangle\langle \alpha|+|\alpha\rangle\langle -\alpha| 
+ |-\alpha\rangle\langle \alpha|),
\label{had}
\end{eqnarray}
considering the projection relations for coherent states $\langle \alpha | \alpha\rangle = 1$ and $\langle \alpha | 
-\alpha\rangle = e^{-2|\alpha|^{2}}$. After this gate operation the state takes the form
\begin{eqnarray}
|\xi_{\pm}\rangle &=& \frac{1}{4}\left(1 + e^{i\phi_{AB}}\right)|-\alpha\rangle \pm \frac{1}{4}\left(1 - e^{i\phi_{AB}}\right)|-\alpha \rangle e^{-2|\alpha|^{2}}\nonumber \\
&-& \frac{1}{4}\left(1 + e^{i\phi_{AB}}\right)|\alpha \rangle e^{-2|\alpha|^{2}} \mp \frac{1}{4}\left(1 - e^{i\phi_{AB}}\right)|\alpha\rangle\nonumber \\
&+& \frac{1}{4}\left(1 + e^{i\phi_{AB}}\right)|\alpha\rangle \pm \frac{1}{4}\left(1 - e^{i\phi_{AB}}\right)|\alpha\rangle e^{-2|\alpha|^{2}}\nonumber \\
&+& \frac{1}{4}\left(1 + e^{i\phi_{AB}}\right)|-\alpha\rangle e^{-2|\alpha|^{2}} \pm \frac{1}{4}\left(1 - e^{i\phi_{AB}}\right)|-\alpha\rangle.
\end{eqnarray} 
Under the condition of negligible overlap, the coherent states $|\alpha\rangle$ and $|-\alpha\rangle$ can be considered orthonormal, 
$\langle \alpha | -\alpha\rangle \approx 0$, being described by a Hilbert space of the 
two-level system spanned by $|\alpha\rangle$ and $|-\alpha\rangle$. The resulting states are 
\begin{eqnarray}
|\xi_{\pm}\rangle &=& \frac{1}{4}\left(1 + e^{i\phi_{AB}}\right)|-\alpha\rangle \mp \frac{1}{4}\left(1 - e^{i\phi_{AB}}\right)|\alpha\rangle \nonumber \\
&+& \frac{1}{4}\left(1 + e^{i\phi_{AB}}\right)|\alpha\rangle \pm \frac{1}{4}\left(1 - e^{i\phi_{AB}}\right)|-\alpha\rangle. 
\end{eqnarray} 
Simplifying for each field state, apart a normalization factor, we have
\begin{eqnarray}
|\psi_{c,+}\rangle = |-\alpha\rangle + e^{i\phi_{AB}}|\alpha\rangle, 
\end{eqnarray}
that corresponds to a selective measurement of the right-handed electron state $|\downarrow\rangle$ and
\begin{eqnarray}
|\psi_{c,-}\rangle = e^{i\phi_{AB}}|-\alpha\rangle + |\alpha\rangle,
\end{eqnarray}
to left-handed electron state $|\uparrow\rangle$. In both cases, the AB phase is transfered to the coherent state superposition.

\section{Bit-enconding of information with AB phases}


One possible use of the generation and control of AB phases is the binary codification. We can propose a bit-enconding in the generation of AB phases by means of the flux control in the solenoid, in particular, we can 
control the imprinting or not of the phase, by absence or not of the flux, due to the switch-on or switch-off of the flux in the solenoid. Since the absence of the gauge potential will lead to absence of the AB phase, the off-state can be characterized by a null vector potential, while 
the on-state can be characterized by its presence. This implies in a bit defined from absence of AB phase, an off-state, and 
the presence of AB phase, the on-state: $bit \rightarrow \lbrace 0, \phi_{AB}\rbrace$.
This can be characterized formally by the binary function dependent of 
presence of a AB phase or its absence:
\begin{eqnarray}
f(\phi_{AB}) = \left\{ \begin{array}{ll}
1 & \mbox{ if } \phi_{AB} \neq {\bf 0}; \\
0 & \mbox{ if } \phi_{AB} \equiv {\bf 0}. \end{array} \right.
\end{eqnarray}
Consequently, a sequential generation of AB phases stands for a sequence of the following 
type $\phi_{AB}^{(0)}\phi_{AB}^{(1)}\phi_{AB}^{(2)}...\phi_{AB}^{(n)}$, where $(i)$ corresponds to the $i$-th order of $i$-th measured phase term. In a string of bits, this results in
\begin{eqnarray}
f_{(0)}(\phi_{AB}^{(0)})f_{(1)}(\phi_{AB}^{(1)})f_{(2)}(\phi_{AB}^{(2)})...f_{(n)}(\phi_{AB}^{(n)}).
\end{eqnarray}
In particular, this codification can be used to implement any ASCII code or a given quantity of bytes. For instance, the string 
of bits $10110$ corresponds to the sequence 
\begin{eqnarray}
& & f_{(0)}(\phi_{AB})f_{(1)}({\bf 0})f_{(2)}(\phi_{AB})f_{(3)}(\phi_{AB})f_{(4)}({\bf 0})\equiv 10110. \label{ufe}
\end{eqnarray}
This binary codification using AB phases could consequently be implemented in any computer device where the presence of the AB effect could be controlled in a integrated way with other components. This could allow any binary message and also contrasts with other possibilities of codification based on chirality or flux orientation. 

\section{Storage of bits enconded by AB phases}
Bits codified by superpositions of states carrying 
AB phases are given by the following state
\begin{eqnarray}
|\prod_{k=0}^{n}f(\phi_{AB}^{(k)}))\rangle &=& \prod_{k=0}^{n}\left(|0\rangle_{(k)} 
+ e^{i\phi_{AB}^{(k)}}|1\rangle_{(k)}\right).
\end{eqnarray}
These states can be stored in appropriate quantum devices, as long as storage times and system interaction be slower than the life time of state. Appropriate unitary evolutions, as in the operator 
\begin{eqnarray}
\hat{\mathcal{N}}=\omega\prod_{k=0}^{n}\hat{n}_{k},
\end{eqnarray}
will leave the codified bit invariant
\begin{eqnarray}
e^{i\hat{\mathcal{N}}t}|\prod_{k=0}^{n}f(\phi_{AB}^{(k)}))\rangle &=& \prod_{k=0}^{n}\left(|0\rangle_{(k)} 
+ e^{i\left(\phi_{AB}^{(k)}+\omega t\right)}|1\rangle_{(k)}\right),
\end{eqnarray}
as long as the evolution time is a period $T_{m}=2m\pi/\omega, m=0,1,...$. As an example, for the state (\ref{ufe}), 
\begin{eqnarray}
e^{i\hat{\mathcal{N}}T_{m}}|10110\rangle &=& |10110\rangle. \label{d1}
\end{eqnarray}
 If we have parallelized high-Q cavities or multimode cavities the transfer can also be implemented to store the state in the form
\begin{eqnarray}
|\prod_{k=0}^{n}f(\phi_{AB}^{(k)}))\rangle &=& \prod_{k=0}^{n}\left(|-\alpha\rangle_{(k)} 
+ e^{i\phi_{AB}^{(k)}}|\alpha\rangle_{(k)}\right),
\end{eqnarray} 
corresponding to a string of bits stored in parallelized high-Q cavities in coherent state superpositions. 

The cavities can be used to store sequentially the bit information with the generation of a product state. 
%
The information enconded in the string of bits can be retransfered to other system or detected by measurement of the interference pattern 
associated to the state. It is important to note that this state can be generated by different methods, imprinting the AB phase in 
parallel AB devices. 
\section{Qudits from enconded states by quantum transfered AB phases}

Due the quantum nature of the system discussed above, the superposition principle will allow superpositions of the form $a|10110\rangle + b|10110\rangle$, as qubits and qudit states. In a general form, the encoded states can be superposed in qudit states
\begin{eqnarray}
|\Psi_{f}\rangle= \sum_{l=0}^{N}s_{l}|\prod_{k=0}^{n}f_{l}(\phi_{AB}^{(k)})\rangle,
\end{eqnarray}
where $\sum_{l=0}^{N}|s_{l}|^{2}=1$ and $f_{l}(\phi_{AB}^{(k)})$ are functions of type $f(\phi_{AB}^{(k)})$. 

This allow the quantum manipulation of these states in a higher hierarchy level. In particular, we have a set of operators 
$\hat{O}$ that can leave these qudits invariant (as the case of an unitary evolution) or implement a quantum gate operation, 
\begin{eqnarray}
|\Psi'_{f}\rangle=\hat{O}|\Psi_{f}\rangle.
\end{eqnarray}
One particular possible application in this context is the use in quantum memories \cite{vetlugin,ding}. The interchanging between 
spintronic and optical scenarios by quantum transfer lead to a more consistent integrability 
among optical and solid state devices in quantum circuits with mixed devices.

\section{One-qubit state modulated by AB phase}

We can also use the quantum transfered AB phase states to codify a qubit of information in eq.(\ref{fincodm}). If after interaction the spin polarization is detected in the state $|\uparrow\rangle$, the non-classical state is projected in a 
superposition of coherent states carrying the AB phase $|\phi_{+}\rangle $. On the other hand, if the measurement 
is realized in the spin polarization $|\downarrow\rangle$, a superposition of coherent states carrying the AB phase terms $|\phi_{-}\rangle$ is 
achieved. Taking the state eq.(\ref{fincodm}) in the normalized form
\begin{eqnarray}
|\phi_{\pm}\rangle = \mathcal{N}_{\pm}\left[\left(1 + e^{i\phi_{AB}}\right)|\alpha'\rangle \pm 
\left(1 - e^{i\phi_{AB}}\right)|\alpha\rangle\right],
\end{eqnarray}
where $\mathcal{N}_{\pm}$ is a normalization factor, the arbitrary coefficients are achieved 
\begin{eqnarray}
|\tilde{\phi}_{\pm}\rangle =  \cos\vartheta_{\pm} |\alpha'\rangle + \sin\vartheta_{\pm}|\alpha\rangle,
\end{eqnarray}
where $\cos^{2}\vartheta_{\pm} +  \sin^{2}\vartheta_{\pm}=1$ and the coefficients depend on the AB phase
\begin{eqnarray}
\vartheta_{\pm}= \tan^{-1}\left(\frac{\pm\left(1 - e^{i\phi_{AB}}\right)}{\left(1 + e^{i\phi_{AB}}\right)}\right).
\end{eqnarray}
This imply that the qubit can be appropriatelly modulated by the AB phase in the quantum transfer.

We can also rewrite the state 
\begin{eqnarray}
|\phi_{\pm}\rangle = 2\mathcal{N}_{\pm}\exp({-i\frac{\phi_{AB}}{2}})\left[\cos(\frac{\phi_{AB}}{2})|\alpha'\rangle \pm 
i\sin(\frac{\phi_{AB}}{2})|\alpha\rangle\right],\label{kl123}
\end{eqnarray} 
We can incorporate the term $2\exp({-i\frac{\phi_{AB}}{2}})$ in the normalization $\mathcal{N}_{\pm}$, such that we now have
\begin{eqnarray}
|\phi_{\pm}\rangle = \mathcal{N}_{\pm}\left[\cos(\frac{\phi_{AB}}{2})|\alpha'\rangle \pm 
i\sin(\frac{\phi_{AB}}{2})|\alpha\rangle\right],\label{kl123}
\end{eqnarray} 
where
\begin{eqnarray}
\mathcal{N}_{\pm}=\frac{1}{\sqrt{1+ Re(\langle \alpha'|\alpha\rangle)\sin(\phi_{AB})}}
\end{eqnarray}
Using the relation for 
\begin{eqnarray}
\langle \alpha'|\alpha\rangle &=& \exp{\left[-\frac{|\alpha'|^{2}+|\alpha|^{2}}{2} + (\alpha')^{*}\alpha\right]},
\end{eqnarray}
that can also be written $\langle \alpha|\alpha'\rangle=\exp{\left[\left(e^{2i\beta t}-1\right)|\alpha|^{2}\right]}$. 

\begin{figure}
\centering
\includegraphics[scale=0.4]{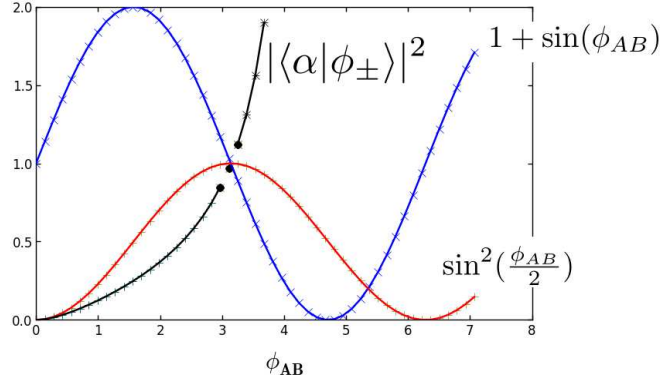}
\caption{(Color online) Plots of $|\langle \alpha|\phi_{\pm}\rangle|^{2}$ (black), $\sin^{2}(\frac{\phi_{AB}}{2})$ (red) and 
$1+ \sin(\phi_{AB})$ (blue). Eq. (\ref{iifal}) has an divergence in $\phi_{AB}=3\pi/2$, after the black dots, that is the case 
where state is purely in $|-\alpha\rangle$ and $\sin\phi_{AB}\neq 2\sin\frac{\phi_{AB}}{2}\cos\frac{\phi_{AB}}{2}$. }
\label{cavity2}
\end{figure}

The state can then be put in the form
\begin{eqnarray}
|\phi_{\pm}\rangle &=& \frac{\cos(\frac{\phi_{AB}}{2})}{\sqrt{1 
+ Re\lbrace\exp{\left[\left(e^{2i\beta t}-1\right)|\alpha|^{2}\right]}\rbrace\sin(\phi_{AB})}}|\alpha'\rangle \nonumber \\
&\pm& i \frac{\sin(\frac{\phi_{AB}}{2})}{\sqrt{1+ Re\lbrace\exp{\left[\left(e^{2i\beta t}
-1\right)|\alpha|^{2}\right]}\rbrace\sin(\phi_{AB})}}|\alpha\rangle, 
\end{eqnarray} 
The projection in the state $|\alpha\rangle$ will lead to
\begin{eqnarray}
\langle \alpha|\phi_{\pm}\rangle = \pm i \frac{\sin(\frac{\phi_{AB}}{2})}{\sqrt{1+ Re\lbrace\exp{\left[\left(e^{2i\beta t}-1\right)|\alpha|^{2}\right]}\rbrace\sin(\phi_{AB})}},
\end{eqnarray} 
Setting the time of interaction $t=\pi/\beta$, adjusting the high-Q cavity parameters, we achieve an AB phase dependent pattern (figure \ref{cavity2}) 
\begin{eqnarray}
|\langle \alpha|\phi_{\pm}\rangle|^{2} = \frac{\sin^{2}(\frac{\phi_{AB}}{2})}{
1+ \sin(\phi_{AB})} \label{iifal}
\end{eqnarray} 
The divergence is in fact the result of

\section{Two-qubit spin-optical state modulated by AB phase}

The state eq. (\ref{c123}) can be rewritten with the time of interaction in $t=\pi/2\beta$, adjusting the high-Q cavity parameters, such that we now have a two qubit state
\begin{eqnarray}
|\psi_{\alpha\alpha'}\rangle &=& a(\phi_{AB})|-\alpha\rangle|\downarrow\rangle 
+ b(\phi_{AB})|\alpha\rangle|\downarrow\rangle \nonumber \\
&+& c(\phi_{AB})|-\alpha\rangle|\uparrow\rangle + d(\phi_{AB})|\alpha\rangle|\uparrow\rangle.
\end{eqnarray}
where the coefficients satisfy the normalization condition
\begin{eqnarray}
|a(\phi_{AB})|^{2} + |b(\phi_{AB})|^{2} + |c(\phi_{AB})|^{2} + |d(\phi_{AB})|^{2} &=& 1, 
\end{eqnarray}
and are given explicitly by
\begin{eqnarray}
a(\phi_{AB})&=& c(\phi_{AB})= 2\mathcal{N}e^{-i\frac{\phi_{AB}}{2}}\cos(\frac{\phi_{AB}}{2}) \\
b(\phi_{AB})&=& -d(\phi_{AB})= 2\mathcal{N}e^{-i\frac{\phi_{AB}}{2}}i\sin(\frac{\phi_{AB}}{2})
\end{eqnarray}
and the normalization factor is 
\begin{eqnarray}
\mathcal{N}=\frac{1}{\sqrt{2\left|1 + e^{i\phi_{AB}}\right|^{2} + 2\left|1 - e^{i\phi_{AB}}\right|^{2}}}=\frac{1}{2\sqrt{2}}
\end{eqnarray} 
We can in fact rewrite the state in the previous form
\begin{eqnarray}
|\psi_{\alpha\alpha'}\rangle &=& \mathcal{N} e^{-i\frac{\phi_{AB}}{2}}\left(\left(e^{-i\frac{\phi_{AB}}{2}} + e^{i\frac{\phi_{AB}}{2}}\right)|-\alpha\rangle|\downarrow\rangle 
-\left(e^{-i\frac{\phi_{AB}}{2}} - e^{i\frac{\phi_{AB}}{2}}\right)|\alpha\rangle|\downarrow\rangle \right) \nonumber \\
&+& \mathcal{N}e^{-i\frac{\phi_{AB}}{2}}\left(\left(e^{-i\frac{\phi_{AB}}{2}} + e^{i\frac{\phi_{AB}}{2}}\right)|-\alpha\rangle|\uparrow\rangle 
+\left(e^{-i\frac{\phi_{AB}}{2}} - e^{i\frac{\phi_{AB}}{2}}\right)|\alpha\rangle|\uparrow\rangle \right),\nonumber \\
\end{eqnarray} 
that simplifies to the state generated by dispersive interaction
\begin{eqnarray}
|\psi_{\alpha\alpha'}\rangle &=& \mathcal{N}\left(\left(1 + e^{i\phi_{AB}}\right)|-\alpha\rangle|\downarrow\rangle 
-\left(1 - e^{i\phi_{AB}}\right)|\alpha\rangle|\downarrow\rangle \right) \nonumber \\
&+& \mathcal{N}\left(\left(1 + e^{i\phi_{AB}}\right)|-\alpha\rangle|\uparrow\rangle 
+\left(1 - e^{i\phi_{AB}}\right)|\alpha\rangle|\uparrow\rangle \right).
\end{eqnarray} 
It is important to note that action of an annihilation operators still leave the state in a two-qubit state
\begin{eqnarray}
|\widetilde{\psi}_{\alpha\alpha'}\rangle=\hat{a}|\psi_{\alpha\alpha'}\rangle, \label{aa}
\end{eqnarray}
where now the coefficients are $\alpha$-dependent
\begin{eqnarray}
|\widetilde{\psi}_{\alpha\alpha'}\rangle &=& \frac{1}{|\alpha|^{2}}\widetilde{a}(\phi_{AB},\alpha)|-\alpha\rangle|\downarrow\rangle 
+ \frac{1}{|\alpha|^{2}}\widetilde{b}(\phi_{AB},\alpha)|\alpha\rangle|\downarrow\rangle\nonumber \\
&+& \frac{1}{|\alpha|^{2}}\widetilde{c}(\phi_{AB},\alpha)|-\alpha\rangle|\uparrow\rangle + \frac{1}{|\alpha|^{2}}\widetilde{d}(\phi_{AB},\alpha)|\alpha\rangle|\uparrow\rangle,
\end{eqnarray} 
where the coefficients satisfy the condition
\begin{eqnarray}
|\widetilde{a}(\phi_{AB},\alpha)|^{2} + |\widetilde{b}(\phi_{AB},\alpha)|^{2} + |\widetilde{c}(\phi_{AB},\alpha)|^{2} + |\widetilde{d}(\phi_{AB},\alpha)|^{2} &=& |\alpha|^{2}. 
\end{eqnarray}
As a consequence, given a coherent state, the AB-phase determines this two-qubit in an hypersphere of radius $|\alpha|$. Obviously, normalization can be obtained with the coefficients
\begin{eqnarray}
\frac{1}{|\alpha|^{2}}\widetilde{a}(\phi_{AB},\alpha)&=& -\frac{\alpha}{|\alpha|^{2}} a(\phi_{AB}), \\
\frac{1}{|\alpha|^{2}}\widetilde{b}(\phi_{AB},\alpha)&=& \frac{\alpha}{|\alpha|^{2}} b(\phi_{AB}), \\
\frac{1}{|\alpha|^{2}}\widetilde{c}(\phi_{AB},\alpha)&=& - \frac{\alpha}{|\alpha|^{2}}c(\phi_{AB}), \\
\frac{1}{|\alpha|^{2}}\widetilde{d}(\phi_{AB},\alpha)&=& \frac{\alpha}{|\alpha|^{2}}d(\phi_{AB}). 
\end{eqnarray}
Another aspect to be emphasized is that the removal of photons by means of the action of annihilation operators can be done without properly destroy the two-qubit state, as what could be the case if the states where just number states. 

We can also analyze the action of a bit-flip operator $\hat{\Pi}$, that will keep the state as a two-qubit state
\begin{eqnarray}
|{\psi}'_{\alpha\alpha'}\rangle=\hat{\Pi}|\psi_{\alpha\alpha'}\rangle, \label{xia}
\end{eqnarray}  
that can also be written by
\begin{eqnarray}
|{\psi}'_{\alpha\alpha'}\rangle &=& a'(\phi_{AB})|-\alpha\rangle|\downarrow\rangle 
+ b'(\phi_{AB})|\alpha\rangle|\downarrow\rangle  \nonumber \\
&+& c'(\phi_{AB})|-\alpha\rangle|\uparrow\rangle + d'(\phi_{AB})|\alpha\rangle|\uparrow\rangle,
\end{eqnarray} 
where now 
\begin{eqnarray}
a'(\phi_{AB}) &=& c(\phi_{AB})=a(\phi_{AB}), \\
b'(\phi_{AB}) &=& d(\phi_{AB})=-b(\phi_{AB}), \\
c'(\phi_{AB}) &=& a(\phi_{AB})=c(\phi_{AB}), \\
d'(\phi_{AB}) &=& b(\phi_{AB})=-d(\phi_{AB}), 
\end{eqnarray}
It is interesting to consider simultaneous action of the operators in (\ref{aa}) and (\ref{xia}). As a consequence, the operator
\begin{eqnarray}
\hat{P}_{\alpha}= \hat{\Pi}\hat{a}\frac{|\alpha|^{2}}{\alpha}
\end{eqnarray}
is a parity operator in the two-qubit spin-optical state
\begin{eqnarray}
\hat{P}_{\alpha}|\psi_{\alpha\alpha'}\rangle=-|\psi_{\alpha\alpha'}\rangle.
\end{eqnarray}
%

\section{Spin-optical density operator modulated by AB phase, dissipation and Lindblad equation}

We now consider the effect of a thermal bath in the density matrix carrying an AB-phase $\hat{\rho}_{\alpha\alpha'}$, corresponding to the 
state $|\psi_{\alpha\alpha'}\rangle$. We start with the dynamics of the density matrix coupled of the system in a thermal bath described by the following Lindblad equation  
\begin{eqnarray}
\frac{d\hat{\rho}_{\alpha\alpha'}}{dt}= -i[\omega_{T} \hat{n},\hat{\rho}_{\alpha\alpha'}] + \left(\hat{\mathcal{L}}(\kappa_{1}\hat{a})+ \hat{\mathcal{L}}(\kappa_{2}\hat{a}^{\dagger})\right)\hat{\rho}_{\alpha\alpha'},
\end{eqnarray}
where $\kappa_{1}=\sqrt{2\kappa (\bar{n}+1)}$, $\kappa_{2}=\sqrt{2\kappa \bar{n}}$ and $\hat{\mathcal{L}}(*)$ is a Lindblad superoperator. 

We also consider the environment effect directly in the state as a result of the action of operators $\hat{\lambda}_{k}(t), k=1,2$
\begin{eqnarray}
\hat{\rho}_{\alpha\alpha'}&=&\hat{\lambda}_{1}(t)|\downarrow \alpha\rangle\langle \downarrow \alpha |\hat{\lambda}_{1}(t)^{\dagger} + \hat{\lambda}_{2}(t)|\uparrow \alpha\rangle\langle \uparrow \alpha |\hat{\lambda}_{2}(t)^{\dagger} \nonumber \\
&+& \hat{\lambda}_{1}(t)|\downarrow \alpha\rangle\langle \uparrow \alpha |\hat{\lambda}_{2}(t)^{\dagger} 
+ \hat{\lambda}_{2}(t)| \uparrow\alpha\rangle\langle \downarrow \alpha |\hat{\lambda}_{1}(t)^{\dagger}.
\end{eqnarray}
Taking into account (\ref{c123}) and the operators encapsulating environment effect, we propose the ansatz
\begin{eqnarray}
\hat{\lambda}_{1}(t) &=& \lambda_{1}(t)\langle\hat{n}\rangle\otimes |\uparrow\rangle\langle \downarrow |\left[\frac{1}{4}\left(1 + e^{i\phi_{AB}}\right)e^{-2i\hat{n}\beta t} -
\frac{1}{4}\left(1 - e^{i\phi_{AB}}\right)\right],  \\
\hat{\lambda}_{2}(t) &=& \lambda_{2}(t)\langle\hat{n}\rangle\otimes |\downarrow\rangle\langle \uparrow |\left[\frac{1}{4}\left(1 + e^{i\phi_{AB}}\right)e^{-2i\hat{n}\beta t} 
+ \frac{1}{4}\left(1 - e^{i\phi_{AB}}\right)\right], 
\end{eqnarray} 
where, for simplicity, a symmetric dissipation effect is considered in the scalar term
\begin{eqnarray}
\lambda_{1}(t)&=& \lambda_{2}(t)=\lambda_{0}(t)(1 + e^{-\Gamma t}),
\end{eqnarray}
$\lambda_{0}(t)$ is a time dependent function from the environment and $\Gamma$ is a coefficient of dissipation to the thermal bath. The operators are rewritten as follows
\begin{eqnarray}
\hat{\lambda}_{1}(t) &=& \lambda_{0}(t)(1 + e^{-\Gamma t})\langle\hat{n}\rangle|\uparrow\rangle\langle \downarrow |\left[\frac{1}{4}\left(1 + e^{i\phi_{AB}}\right)e^{-2i\hat{n}\beta t} -
\frac{1}{4}\left(1 - e^{i\phi_{AB}}\right)\right], \label{opk0} \\
\hat{\lambda}_{2}(t) &=& \lambda_{0}(t)(1 + e^{-\Gamma t})\langle\hat{n}\rangle|\downarrow\rangle\langle \uparrow |\left[\frac{1}{4}\left(1 + e^{i\phi_{AB}}\right)e^{-2i\hat{n}\beta t} 
+ \frac{1}{4}\left(1 - e^{i\phi_{AB}}\right)\right]. \label{opk1}
\end{eqnarray} 
Under explicit action of (\ref{opk0}) and (\ref{opk1}) on the coherent state, these operators are related to spin flipping operators 
\begin{eqnarray}
\hat{\lambda}_{1}(t) |\alpha\rangle &=& \frac{1}{4}\left[\left(1 + e^{i\phi_{AB}}\right)|e^{-2i\beta t}\alpha\rangle 
-\left(1 - e^{i\phi_{AB}}\right)|\alpha\rangle \right]\lambda_{0}(t)(1 + e^{-\Gamma t})\langle\hat{n}\rangle|\uparrow\rangle\langle \downarrow |,\nonumber \\
&& \\
\hat{\lambda}_{2}(t)|\alpha\rangle &=& \frac{1}{4}\left[\left(1 + e^{i\phi_{AB}}\right)|e^{-2i\beta t}\alpha\rangle 
+\left(1 - e^{i\phi_{AB}}\right)|\alpha\rangle \right] \lambda_{0}(t)(1 + e^{-\Gamma t})\langle\hat{n}\rangle|\downarrow\rangle\langle \uparrow |. \nonumber \\
\end{eqnarray}
The density operator evolves according to the time evolution of these operators 
\begin{eqnarray}
\frac{d\hat{\rho}_{\alpha\alpha'}}{dt}&=& \frac{d\hat{\lambda}_{1}(t)}{dt}|\downarrow \alpha\rangle\langle \downarrow \alpha |\hat{\lambda}_{1}(t)^{\dagger} + \hat{\lambda}_{1}(t)|\downarrow \alpha\rangle\langle \downarrow \alpha |\frac{d\hat{\lambda}_{1}(t)^{\dagger}}{dt} \nonumber \\
&+& \frac{d\hat{\lambda}_{2}(t)}{dt}|\uparrow \alpha\rangle\langle \uparrow \alpha |\hat{\lambda}_{2}(t)^{\dagger} + \hat{\lambda}_{2}(t)|\uparrow \alpha\rangle\langle \uparrow \alpha |\frac{d\hat{\lambda}_{2}(t)^{\dagger}}{dt} \nonumber \\
&+& \frac{d\hat{\lambda}_{1}(t)}{dt}|\downarrow \alpha\rangle\langle \uparrow \alpha |\hat{\lambda}_{2}(t)^{\dagger} + \hat{\lambda}_{1}(t)|\downarrow \alpha\rangle\langle \uparrow \alpha |\frac{d\hat{\lambda}_{2}(t)^{\dagger}}{dt} \nonumber \\
&+& \frac{d\hat{\lambda}_{2}(t)}{dt}|\uparrow \alpha\rangle\langle \downarrow \alpha |\hat{\lambda}_{1}(t)^{\dagger} + \hat{\lambda}_{2}(t)|\uparrow \alpha\rangle\langle \downarrow \alpha |\frac{d\hat{\lambda}_{1}(t)^{\dagger}}{dt}. 
\end{eqnarray}
This time dependence can also be rewritten
\begin{eqnarray}
\frac{d\hat{\rho}_{\alpha\alpha'}}{dt}&=& \frac{d(\hat{\lambda}_{1}(t)|\alpha\rangle)}{dt}|\downarrow\rangle\langle \downarrow|(\langle\alpha |\hat{\lambda}_{1}(t)^{\dagger}) + (\hat{\lambda}_{1}(t)|\alpha\rangle)|\downarrow\rangle\langle \downarrow|\frac{d(\langle\alpha |\hat{\lambda}_{1}(t)^{\dagger})}{dt} \nonumber \\
&+& \frac{d(\hat{\lambda}_{2}(t)|\alpha\rangle)}{dt}|\uparrow\rangle\langle \uparrow|(\langle\alpha |\hat{\lambda}_{2}(t)^{\dagger}) + (\hat{\lambda}_{2}(t)|\alpha\rangle)|\uparrow\rangle\langle \uparrow|\frac{d(\langle\alpha |\hat{\lambda}_{2}(t)^{\dagger})}{dt} \nonumber \\
&+& \frac{d(\hat{\lambda}_{1}(t)|\alpha\rangle)}{dt}|\downarrow\rangle\langle \uparrow|(\langle\alpha |\hat{\lambda}_{2}(t)^{\dagger}) + (\hat{\lambda}_{1}(t)|\alpha\rangle)|\downarrow\rangle\langle \uparrow|\frac{d(\langle\alpha |\hat{\lambda}_{2}(t)^{\dagger})}{dt} \nonumber \\
&+& \frac{d(\hat{\lambda}_{2}(t)|\alpha\rangle)}{dt}|\uparrow\rangle\langle \downarrow|(\langle\alpha |\hat{\lambda}_{1}(t)^{\dagger}) + (\hat{\lambda}_{2}(t)|\alpha\rangle)|\uparrow\rangle\langle \downarrow|\frac{d(\langle\alpha |\hat{\lambda}_{1}(t)^{\dagger})}{dt} 
\end{eqnarray}
Since the action of the annihilation operators will give a phase contribution $e^{-2i\beta t}$ in $\alpha'$, we have
\begin{eqnarray}
\hat{a}\hat{\lambda}_{1}(t) |\alpha\rangle &=& \frac{1}{4}\left[\left(1 + e^{i\phi_{AB}}\right)\alpha'|\alpha'\rangle 
-\left(1 - e^{i\phi_{AB}}\right)\alpha|\alpha\rangle \right]\lambda_{0}(t)(1 + e^{-\Gamma t})\langle\hat{n}\rangle|\uparrow\rangle\langle \downarrow | \nonumber \\
&&  \\
\hat{a}\hat{\lambda}_{2}(t)|\alpha\rangle &=& \frac{1}{4}\left[\left(1 + e^{i\phi_{AB}}\right)\alpha'|\alpha'\rangle 
+\left(1 - e^{i\phi_{AB}}\right)\alpha|\alpha\rangle \right] \lambda_{0}(t)(1 + e^{-\Gamma t})\langle\hat{n}\rangle|\downarrow\rangle\langle \uparrow |\nonumber \\
\end{eqnarray}
\begin{figure}
\centering
\includegraphics[scale=0.5]{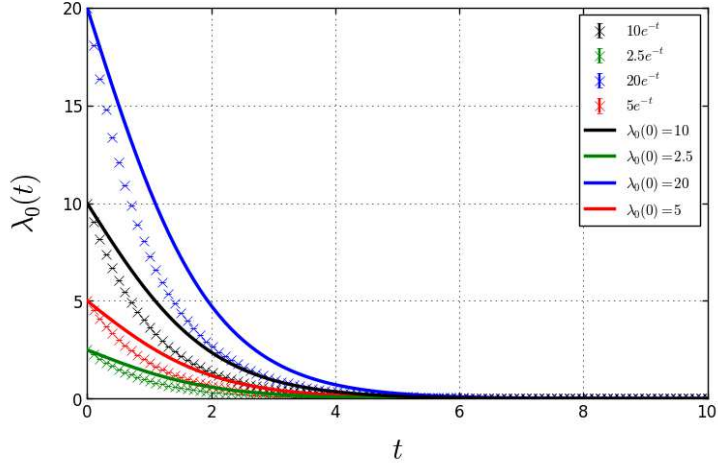}
\caption{(Color online) $\lambda_{0}(t)$ for different initial values, compared to an exponential decay.}
\label{cavity3}
\end{figure}
In a low temperature regime, $\bar{n}_{th}\approx 0$, $\langle \hat{n} \rangle\approx |\alpha|^{2}$ and $-i[\omega_{T} \hat{n},\hat{\rho}_{\alpha\alpha'}]\approx -i[\omega_{T} \langle \hat{n} \rangle,\hat{\rho}_{\alpha\alpha'}]$, such that the Lindblad operators simplify
\begin{eqnarray}
\hat{\mathcal{L}}(\kappa_{1}\hat{a})\hat{\rho}_{\alpha\alpha'} &=& \kappa_{1}^{2}\left[\hat{a}\hat{\rho}_{\alpha\alpha'}\hat{a}^{\dagger} 
-|\alpha|^{2} \hat{\rho}_{\alpha\alpha'}\right], \\
\hat{\mathcal{L}}(\kappa_{2}\hat{a}^{\dagger})\hat{\rho}_{\alpha\alpha'} &=& \kappa_{2}^{2}\left[\hat{a}^{\dagger}\hat{\rho}_{\alpha\alpha'}\hat{a}
-|\alpha|^{2} \hat{\rho}_{\alpha\alpha'}\right].
\end{eqnarray}
We also consider the spin interaction and the dissipation effect occur along different time regimes, for appropriate spin interaction $\alpha'\approx \alpha$, the state is left
\begin{eqnarray}
\hat{a}\hat{\lambda}_{1}(t) |\alpha\rangle &=& \frac{1}{4}\left[\left(1 + e^{i\phi_{AB}}\right)\alpha|\alpha\rangle 
-\left(1 - e^{i\phi_{AB}}\right)\alpha|\alpha\rangle \right]\lambda_{0}(t)(1 + e^{-\Gamma t})\langle\hat{n}\rangle|\uparrow\rangle\langle \downarrow |\nonumber \\
\hat{a}\hat{\lambda}_{2}(t)|\alpha\rangle &=& \frac{1}{4}\left[\left(1 + e^{i\phi_{AB}}\right)\alpha|\alpha\rangle 
+\left(1 - e^{i\phi_{AB}}\right)\alpha|\alpha\rangle \right] \lambda_{0}(t)(1 + e^{-\Gamma t})\langle\hat{n}\rangle|\downarrow\rangle\langle \uparrow |\nonumber 
\end{eqnarray}
Under this regime the first Lindblad operator vanishes $\hat{\mathcal{L}}(\kappa_{1}\hat{a})\hat{\rho}_{\alpha\alpha'}\approx 0$, leaving the Lindblad equation in the following form
\begin{eqnarray}
\frac{d\hat{\rho}_{\alpha\alpha'}}{dt}= \kappa_{2}^{2}\left[\hat{a}^{\dagger}\hat{\rho}_{\alpha\alpha'}\hat{a}
-|\alpha|^{2} \hat{\rho}_{\alpha\alpha'}\right].
\end{eqnarray}
We also consider that, under dissipation, the coherent state will be brought to a number state $| n_{\alpha} \rangle$, $|\alpha| > n_{\alpha}\gg 1$, such that
\begin{eqnarray}
\hat{a}^{\dagger}\hat{\lambda}_{1}(t)|n_{\alpha}\rangle &\approx& \frac{1}{2}\left[e^{i\phi_{AB}}\sqrt{n_{\alpha}+1}|n_{\alpha}+1\rangle \right]\lambda_{0}(t)(1 + e^{-\Gamma t})|\alpha|^{2}|\uparrow\rangle\langle \downarrow | \\
\hat{a}^{\dagger}\hat{\lambda}_{2}(t)|n_{\alpha}\rangle &\approx& \frac{1}{2}\left[\sqrt{n_{\alpha}+1}|n_{\alpha}+1\rangle 
 \right] \lambda_{0}(t)(1 + e^{-\Gamma t})|\alpha|^{2}|\downarrow\rangle\langle \uparrow |
\end{eqnarray}
and we can finally write
\begin{eqnarray}
\hat{a}^{\dagger}\hat{\lambda}_{1}(t)|n_{\alpha}\rangle &\approx& \frac{1}{2}\left[e^{i\phi_{AB}}\sqrt{n_{\alpha}}|n_{\alpha}\rangle \right]\lambda_{0}(t)(1 + e^{-\Gamma t})|\alpha|^{2}|\uparrow\rangle\langle \downarrow | \\
\hat{a}^{\dagger}\hat{\lambda}_{2}(t)|n_{\alpha}\rangle &\approx& \frac{1}{2}\left[\sqrt{n_{\alpha}}|n_{\alpha}\rangle 
 \right] \lambda_{0}(t)(1 + e^{-\Gamma t})|\alpha|^{2}|\downarrow\rangle\langle \uparrow | \\
\hat{\lambda}_{1}(t)|n_{\alpha}\rangle &\approx& \frac{1}{2}\left[e^{i\phi_{AB}}|n_{\alpha}\rangle \right]\lambda_{0}(t)(1 + e^{-\Gamma t})|\alpha|^{2}|\uparrow\rangle\langle \downarrow | \\
\hat{\lambda}_{2}(t)|n_{\alpha}\rangle &\approx& \frac{1}{2}\left[|n_{\alpha}\rangle 
 \right] \lambda_{0}(t)(1 + e^{-\Gamma t})|\alpha|^{2}|\downarrow\rangle\langle \uparrow | 
\end{eqnarray}
The Lindblad equation in its reduced form is essentially 
the scalar differential equation 
\begin{eqnarray}
\frac{d (\lambda_{0}(t)^{2}(1 + e^{-\Gamma t})^{2})}{dt}&=& \frac{-\kappa_{2}^{2}}{4}\left(|\alpha|^{2} - n_{\alpha}^{2}\right)\lambda_{0}(t)^{2}(1 + e^{-\Gamma t})^{2},
\end{eqnarray}
whose solution is given by (figure \ref{cavity3})
\begin{eqnarray}
\lambda_{0}(t) &=& \frac{2\lambda_{0}(0)}{(1 + e^{-\Gamma t})}e^{-\frac{1}{2}\kappa_{2}^{2}\left(|\alpha|^{2}-n_{\alpha}^{2}\right)t},
\end{eqnarray}
and a density operator solution 
\begin{eqnarray}
\hat{\rho}_{\alpha\alpha'}(t)&=&\frac{1}{4}\lambda_{0}(t)^{2}
(1 + e^{-\Gamma t})^{2}|\alpha|^{4}|n_{\alpha}\rangle\langle n_{\alpha}|\left(|\uparrow\rangle\langle \uparrow| + |\downarrow\rangle\langle \downarrow|\right) \nonumber \\
&+& \frac{1}{4}\lambda_{0}(t)^{2}
(1 + e^{-\Gamma t})^{2}|\alpha|^{4}|n_{\alpha}\rangle\langle n_{\alpha}|\left(e^{i\phi_{AB}}|\uparrow\rangle\langle \downarrow| + e^{-i\phi_{AB}}|\downarrow\rangle\langle \uparrow|\right),
\end{eqnarray}


\section{Conclusions}

We proposed a quantum transfer of Aharonov-Bohm phase to a non-classical state of light from a 
TST, a spintronic device, to a high-Q cavity field, a quantum optical device. The quantum state generated in the 
TST is injected in the cavity in a superposed spin state with the presence of an AB phase. The interaction with the 
coherent state then leads to an entanglement. A measurement of the spin state finally projects the AB phase in the non-classical state of light, whose 
quantum optical manipulation leads to a QST. We also discussed the implementation of this proposal in bit encondings involving quantum system and 
qudit states associated to quantum memories.  The quantum transfer used to realize the interchanging among 
spintronic and optical systems imply in a more consistent integrability in quantum circuits involving mixed devices. 

The presence of a phase in a quantum state can be used for the generation of quantum interference effects that can be quantum 
transfered from a solid state to quantum optical state. The transfer of an AB phase to a quantum optical device also leads to the possibility of transfer 
the AB interferences from the solid state to the optical device. Bit and qubit condifications can be realized by modulation of the AB phase. In particular, 
a qubit state can be generated with the coefficients modulated by the AB phase in the quantum transfer.

We also explored the presence of dissipative effects, in particular, exploring the Lindblad equation, considering the system in a thermal bath. A particular Lindblad equation was then achieved and solved for the system in the presence of an AB phase. 

Given the experimental advances in the generation and control of AB effect, such a quantum transfer can be experimentally realized and further progress 
can be achieved. Our proposal can also be useful to make progress in methods of quantum information associated to 
modern techniques in synthetic gauge fields. Taking advantange of quantum information methods for control and manipulation of quantum interference phenomena, the methods of synthetic gauge fields can be improved and spintronics can be integrated with quantum optical devices.

\section{Acknowledgements}
The author thanks the support by projects FAPEMA (Brazil)- APCINTER-00273/14, Enxoval – UFMA PPPG N03/2014 (Brazil), institutionalized by UFMA-Res.No1342-CONSEPE-Art1-III-1150/2015-33, 
UFMA-Res.No1342-CONSEPE Art1-IV-1151/2015-88 and FAPEMA-UNIVERSAL-01401/16.


\begin{thebibliography}{99}
\bibitem{aharo1} Y. Aharonov, D. Bohm, Phys. Rev. {\bf 115} (1959) 485.
\bibitem{aharo2} Y. Aharonov, D. Bohm, Phys. Rev. {\bf 123} (1961) 1511.
\bibitem{olariu3} S. Olariu, I.I. Popescu, Rev. Mod. Phys. {\bf 57} (1985) 339.
\bibitem{caprez4} A. Caprez, B. Barwick, H. Batelaan, Phys. Rev. Lett. {\bf 99} (2007) 210401.
\bibitem{hernandez5} A. R. Hern\'andez, C. H. Lewenkopf, Phys. Rev. Lett. {\bf 103} (2009) 166801.
\bibitem{vaidman6} L. Vaidman, Phys. Rev. A {\bf 86} (2012) 040101(R).
\bibitem{duca} L. Duca, et al., Science {\bf 347} (2015) 288.
\bibitem{niu} Z.P. Niu, Eur. Phys. J. B {\bf 82} (2011) 153.
\bibitem{charlier} J-C. Charlier, X. Blase, S. Roche, Rev. Mod. Phys. {\bf 79} (2007) 677.
\bibitem{edgcombe} C. J. Edgcombe, J. C. Loudon J. Phys.: Conf. Ser. {\bf 371} (2012) 012006.
\bibitem{nitta} J. Nitta, T. Koga, H. Takayanagi, Physica E {\bf 12} (2002) 753.
\bibitem{splet} J. Splettstoesser, M. Moskalets, M. B\"uttiker, Phys. Rev. Lett. {\bf 103} (2009) 076804.
\bibitem{eckle} H.-P. Eckle, H. Johannesson, C. A. Stafford, Phys. Rev. Lett. {\bf 87} (2001) 016602.
\bibitem{matityahu} S. Matityahu, A. Aharony, O. Entin-Wohlman, S. Tarucha, New J. Phys. {\bf 15} (2013) 125017. 
\bibitem{li} E. Li, B. J. Eggleton, K. Fang, S. Fan, Nature Comm. {\bf 5} (2014) 3225.
\bibitem{sigurdsson} H. Sigurdsson, O.V. Kibis, I.A. Shelykh, Phys. Rev. B {\bf 90} (2014) 235413. 
\bibitem{alexeev} A. M. Alexeev, I. A. Shelykh, M. E. Portnoi, Phys. Rev. B {\bf 88} (2013) 085429. 
\bibitem{alexeev2} A. M. Alexeev and M. E. Portnoi, Phys. Rev. B {\bf 85} (2012) 245419.
\bibitem{ramaglia} V. M. Ramaglia, F. Ventriglia, G. P. Zuchelli, Phys. Rev. B {\bf 52} (1995) 8372.
\bibitem{vieira} M. Vieira, A. M. M. Carvalho, C. Furtado, Phys. Rev. A {\bf 90} (2014) 012105.
\bibitem{dalibard} J. Dalibard, F. Gerbier, G. Juzeliunas, P. Ohberg, Rev. Mod. Phys. {\bf 83} (2011) 1523.
\bibitem{mironova} P. V. Mironova, M. A. Efremov, W. P. Schleich, Phys. Rev. A {\bf 87} (2013) 013627.
\bibitem{casher} Y. Aharonov and A. Casher, Phys. Rev. Lett. {\bf 53} (1984) 319.
\bibitem{andrade2} F. M. Andrade, E. O. Silva, T. Prudencio, C. Filgueiras, J. Phys. G: Nucl. Part. Phys. {\bf 40} (2013) 075007.  
\bibitem{kovalev} A. A. Kovalev, et al., Phys. Rev. B {\bf 76} (2007) 125307. 
\bibitem{hartnoll} S. A. Hartnoll, Phys. Rev. Lett. {\bf 98} (2007) 111601.
\bibitem{andrade3} E. O. Silva, F. M. Andrade, H. Belich, C. Filgueiras, Eur. Phys. J. C {\bf 73} (2013) 2402.
\bibitem{bakke} K. Bakke, E. O. Silva, H. Belich, J. Phys. G: Nucl. Part. Phys. {\bf 39} (2012) 055004. 
\bibitem{ding} F. Ding, et al. Phys. Rev. B {\bf 82} (2010) 075309.
\bibitem{maciejko} J. Maciejko, E-A. Kim, X-L. Qi, Phys. Rev. B {\bf 82} (2010) 195409.
\bibitem{muller} K. M\"uller, et al. Phys. Rev. Lett. {\bf 114} (2015) 233601. 
\bibitem{brune} A. Roth, C. Br\"une, H. Burhmann, L. W. Molenkamp, J. Maciejko, X-L. Qi, S-C. Zhang, Science {\bf 318} (2009) 294.
\bibitem{liberto} M. Di Liberto et al. Nature Communications {\bf 5} (2014) 5735.
\bibitem{prudencio} T. Prudencio, Int. J. Quantum Inf. {\bf 11} (2013) 1350024.
\bibitem{das} S. Datta, B. Das, Appl. Phys. Lett. {\bf 56} (1990) 665.
\bibitem{zutic} I. Zutic, J. Fabian, S. Das Sarma, Rev. Mod. Phys. {\bf 76} (2004) 323.
\bibitem{vetlugin} A. N. Vetlugin, I. V. Sokolov, Eur. Phys. J. D {\bf 68} (2014) 269.
\bibitem{ding} D-S. Sheng, W. Zhang, Z-Y. Zou, S. Shi, B-S. Shi, G-C. Guo, Nature Photonics {\bf 9} (2015) 332.
\end{thebibliography}
\end{document}